# Digitalized Responsive Logical Interface Application

Maxwell Scale Uwadia Osagie[1], Kingsley O. Obahiagbon[1],
Amanda Iriagbonse Inyang[1], Joy Amenze Osagie[2]

[1]Department of Mathematics and Computer Science, Faculty of Basic and Applied Sciences Benson Idahosa University, G.R.A., Benin City, Edo State, Nigeria
[2]M.Sc Student: Department of Computer Science, School of Post Graduate Studies, Benson Idahosa University, G.R.A., Edo State, Benin City, Nigeria

*Abstract:* Examination and record unit (ERU) is one of the most sensitive units in Nigeria universities systems. Most data kept in the domain of the ERU are usually for decisions and onward review. Hence, the preservation of ERU data formed one of the focal point of institution sustainability. Over the years, these data have one way or the other lost it integrity and as a result resulting to lack of trust in its output due to the method used in safeguarding student's data. Though, this section in tertiary institutions is yet to be given adequate considerations, we hope to use this medium to not only draw the Nigeria tertiary institutions to what become of the institution if this current existing method is not improved upon but to made know the method used by most advanced country tertiary institutions that have so far demonstrated strong base in its data preservation. However, non logical interfaces of systems synchronized have over the years characterized the operational functions of the Nigeria tertiary Examination and Records Unit (ERU). The quest for proper protection of data in the ERU and its accessibility led to the design of a Digitalized Responsive Logical Interface Application (DRLIA) with an embedded feature such as keeping ERU data safer and accessible at any given point. The system works in three modules [i] Password/Staff number Synchronization, [ii] Encryption (token) Synchronization and [iii] i and ii synchronizations. This method help prevent unauthorized users from gaining access to the stored data, it also has the capability of recording time the authorized user gained access to the system in case of system theft. It has the ability to render the data unreachable, thereby making it not useful.

## 1. INTRODUCTION

A Digitalized Responsive Logical Interface (DRLI) is an application interface that utilizes the coherent principle of 0's and 1's in implementing data/information authentication, thereby preventing unauthorized persons from gaining access to information meant for decision making. Information as a variable tool play a major role in bridging geographical boundaries amongst countries, and its inherent attributes allows decision making and also bring about continuous flow of data/information amongst people. Since the introduction of computer to reduce human complexity it has experienced a tremendous turnaround over the years [1].  It will be a welcome idea if organizations/countries that are physically and environmentally apart be made to know better authentication modules that will further increasing the velocity of data/information manipulation (computation).  Information is use in variety of things [2]. From the sources of gathering, processes and the output which serves as decision making remains a vital part of today changing world. Countries and organizations have employed different strategies in ensuring data safety[3], the more finances pumped into data/information security by concerned countries and individual around the globe the more vulnerable it has become due to conflict of interest.  Ensuring the safety of data/information remains the sole responsibility of the offices saddled with the responsibility of keeping and retrieving it (data/information) for onward processing[5],[6]. It is not out of point to say that most personnel who worked in these offices are either intimidated, bought or many at times payback time for the organization they represent by allowing sensitive data/information to be seen in public domain and thereafter making such data/information not useful for the original purpose, or simply put they lack the technical ability to keep data safe

Interface development is a systems engineering activity because of the basic needs to break down large problems into many related smaller modules/components and the smaller problems have interfaces between them that must be compatible across all terminals [7]. Interface is a point at which independent systems or components meet and act or communicates[8],[9]. Interfaces can exist between system elements and environment. Interfaces can enhance system capability but a trade-off must be made between internal and external complexity. In software engineering, it is recognized to be an attempt to balance the concepts of cohesion and coupling[10]. Interfaces between hardware and software must be carefully examined due to its impact on the complexity of software intensiveness. Interface represents the beauty of any application because through the interface, access is gained into the main application[11], [12],[13].

The quest to ensuring adequate secure of data, translates to the design of a digitalized responsive logical interface application with state of the art security apparatus such as username, password and encryption techniques in preventing intruders or hackers from manipulating such application in order to steal valuable information. To every interface comes module of different natures[14], just like a well structure design of a home where the outside of the building tell more of what might be embedded. Interface does to an application or computer system what entrance gate does to a building. The nature of a building (house) determines how beautiful and solid the entrance gate will be. Applications meant for sensitive materials or documents should be well protected by a well structured interface. Having the understanding that no document or system is secured enough, some establishments are employing different security measure in keeping useful data from non useful data [15].The purpose for data security cannot be overemphasis, systems that are homogenous and heterogeneous share resources for different reasons and this help reduce the overhead cost of what organizations and companies ought to spend if these data were to be shared locally (manually).

The quest to secure data for future use has made many establishments and companies engage in seeking services of data security expert. The use of strong and well structure interfaces with encryption have creates confidence on the part of the system users irrespective of the challenges surrounding the integrity of computer systems. [16]

*Words and meaning*

- **Digitalized:**  Information/data converted into digital format
- **Responsive:** Reacting quickly and in positive way
- **Logical:** Reasonable and sensible
- **Interface:** The way computer receive  and present information to user
- **Application:** A system/package designed for particular purpose

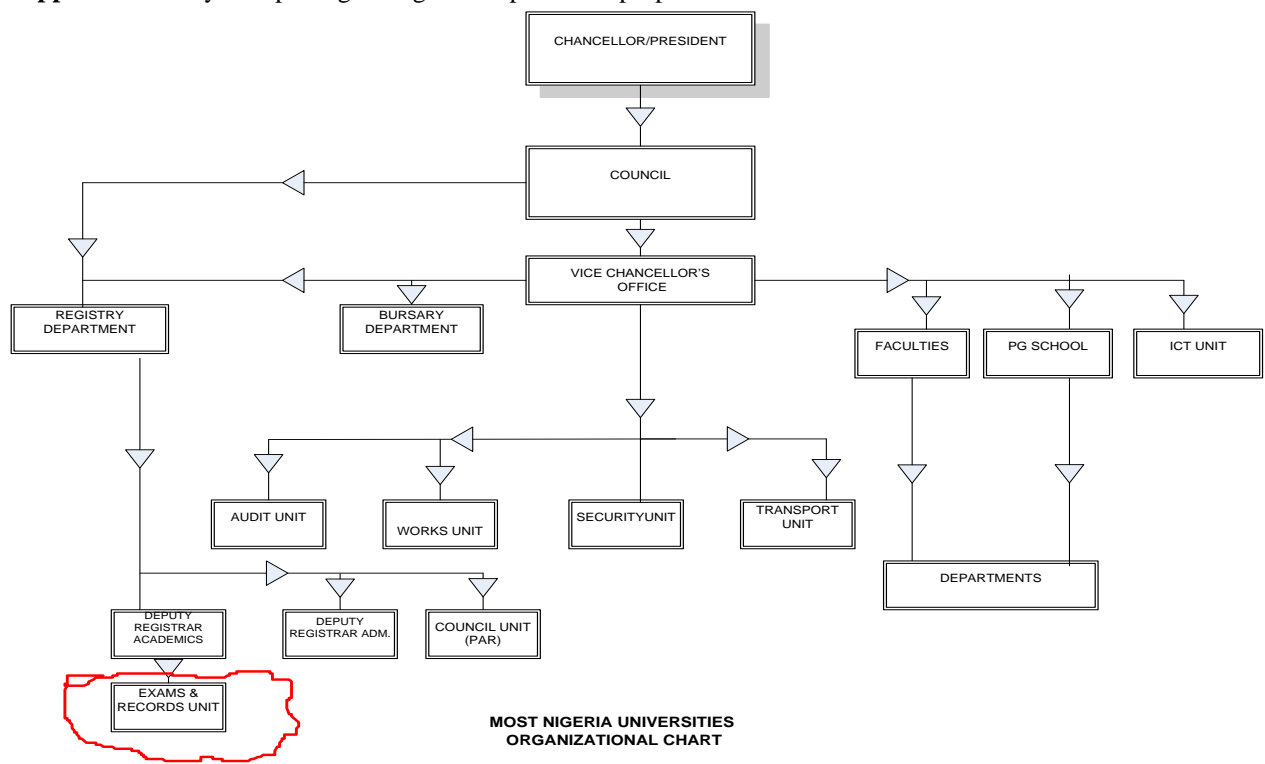

**Figure 1: Organizational Chart of Most Nigeria Universities**

## II. THE ROLE OF EXAMINATION AND RECORD UNIT IN A UNIVERSITY

A close examination to figure 1 shows different structures existing within the Nigeria universities and of great importance to this paper is the red session of the chart which profoundly unveil the Examination and Record Unit (ERU). The registry unit of any Nigeria University is headed by a Registrar who is often refer to as the second most ranked principle officers; he is also the secretary to the university board of council. He presides over documents issued out internally and externally. He is directly answerable to the vice chancellor of the university and his job jurisdiction cut across all aspects of the university as demonstrated by figure1. Every administrative staff in the university i.e. those not under the vice chancellors office falls under his office. The registry headed by a Registrar reserves the right to move administrative staff from one office to another as approved by the vice chancellor and this is generally called **cabinet readjustment.** The registry as a department has sub units or offices that report to the Registrar either by writing or verbal on daily basics. One of these sub units is the Examinations and Records Unit

*Examinations and Record Unit Activities:*

From the above analysis it is a well know fact that the examinations and records unit of Nigeria Universities report to the Registrar on their daily activities. Their duty include the following

   i. Act as the middle man between the university and students in result matter
  ii. Processes each level of student result as approve by senate
 iii. Does the cumulative result in collaboration with the faculties and departments that is eventually issue to final year students
  iv. Responsible in generating student's transcript
   v. Responsible in receive transcript from accredited academic institutions
  vi. Responsible in sending student's transcript to other institutions on request
 vii. Store/keep student's results for future use
viii. Does the verification of student entry result in collaboration with the committee setup by the university management

*Examination and Record Unit Challenges:*

A retrospect into the activities of the Examinations and Records Unit gives a better understanding of the enormous task the unit is saddle with. The biggest challenge to the present method of operation is the vulnerability of sensitive data due to lack of proper security measure. Security remains a major problem to human race and billions of dollars are been spent around the globe in bringing about peace. Academic institutions are not immune from this security challenges, same also are Nigeria Universities. As the management and staff of academic institutions tries to be vigilant in their operations, students who are under their tutelage capitalize on the pitfall of these staff in committing illicit act. This was seen from the survey conducted. From the survey it was discovered that the method currently being operated failed to meet the current trend of modern day computing. The operation of having a student result stored in a file cabinet or ordinary computer memory will continue to bring about inefficiency amongst operational staff and thereafter, making them unproductive.

## III. RELATED WORK

Keeping data out of reach from intruders has continued to be key research area amongst information communication technology expert and researchers due to the sensitive nature of the information kept for onward review[17], [18]. Research has shown a turnaround in the manner in which organizations approach security problem and some have resulted to a migration from the normal method to a more profound approach by simply employing third party. The use of cloud has gained wider interest in recent times [19]. In cloud computing rather than invest on gadgets, organization could subscribe to services that will best serve the interest of the organization in data security. In cloud computing software as a service (SaS), platform as a service (PaS) and infrastructure as service (IaS) work as one to ensure optimal delivering of smooth service delivering to users [20].  There are many things that could make organization such as universities afraid when dealing with sensitive information and the following are security threat that may occur if adequate steps are not adopted to preserve data within a predefined system structure.

  i. Vulnerability of data

  ii. Violation of data purpose

  iii. Camouflaged operations on the part of the data handlers

  iv. Denial of service

  v. Wrong use of data

  vi. Threats such as inbound and outbound threat

  vii. Access control

  viii. Distributed Denial of service (DDoS)

  ix. Cyber terrorism

  x. Phishing

  xi. Etc.

Organizations in the $21^{st}$ century understand the capacity of both valid and invalid data but in this part of the world low awareness has be the issue of sustainability. Data security is the act of ensuring data protection from both intruders and hackers. With the current development is software tools capable of bridging the integrity of data[21], organization such as academic institution should ensure to adopt a better security method that could prevent and preserve data meant for onward review. According to [22], the state of cybercrime among youths has not only skyrocketed in recent times but has become more worrisome. The sustainability of data must be a responsibility of all if Nigeria as a country must join the league of developed countries.

## IV.   DESIGNED WORK IN PROSPECT

The proposed system if fully implemented will resolve the laid down issues associated with the old system. The system created is web base enabled and has the following features/modules

  i. Breaking the entry authentication into three modules (such as [i] Password/Staff number Synchronization, [ii] Encryption Synchronization and [iii] i and ii Synchronizations

  ii. Prevent unauthorized users of gaining access to such data/information

  iii. Has the capability of recording time the authorized user gained access to the system?

  iv. In case of system theft render the data unreachable, thereby making it not useful

  v. Accessing data on a real time

  vi. Flexibility in the part of the authorized user

The above modules were reached after series of investigations and information gathered in ascertaining the level of efficiency and infectiveness of the current operational role of examinations and records unit existing system.

The propose system will not only eradicate security problem that is currently associated with the present system but will also reduce redundancy amongst operational staff and data, thereby returning money on investment and increase the growth rate of the university since the unit bridge the gap between management and students following the sensitive nature of the office.

## V.   SYSTEM DESIGN

The system design uses Hypertext Markup Language, Hypertext Processor (PHP), CSS and MYSQL as a database to implement the components and modules synchronization. The existing system also known as the present system before the introduction of the proposed system lacks several tools and functions. The vital issues outline in the course of the analysis showed the deficiency of the existing system and should urgently be replaced with the recommendation of this paper. The quest to ensure trending in technology amongst agencies, educational intuitions led to the design and development of this system.

The system design employed two security measures in bringing about trust between operational staff. The design structure bellow showed the various components that form the entire system. The system design has the following mechanism that facilitate heterogeneous and homogenous access of the system

*Preliminary Design*

Interface to any system represent the entry point to which access is made. Organizations and institutions hardly pay more attention to this part due to reasons best know to them. In the course of the design, many ideas translated to the features to which this proposed system is built. To ensure constant usage of this system, adequate care must be taken in ensuring that the following are kelp to date

   i. The database (repository) gadget
  ii. The machine (the hardware)
 iii. Training

*Advantages of the Proposed System:*

   i. Full implementation of this system proposed will help the management of Nigeria Tertiary Institutions (Universities) become one of the fewest institutions who has made the examination and record unit accessible to the comfort of staff's home within Africa academic Institution thereby making them be part of the developed countries
  ii. Prevent unauthorized users from gaining access to such data/information
 iii. Has the capability of recording time the authorized user gained access to the system?
  iv. In case of system theft render the data unreachable, thereby making it not useful
   v. Accessing data on a real time
  vi. Flexibility  the authorized users

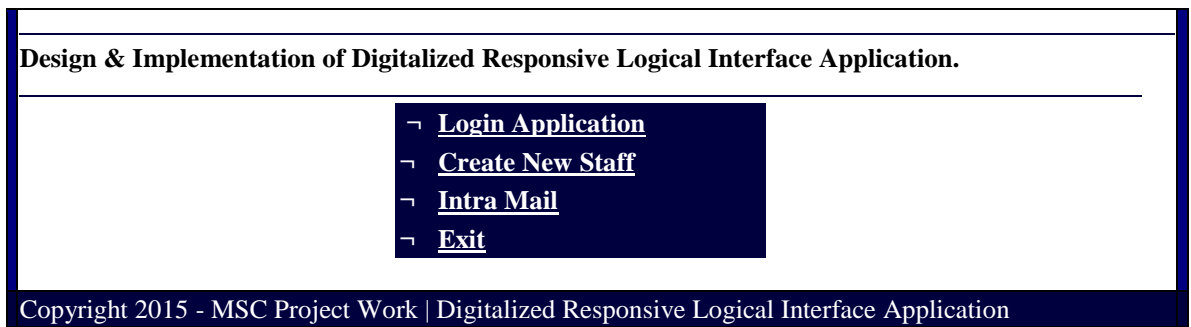

**Figure 2: Home page**

The figure 2 servers as the entering point to both existing users and new users. The different modules display within the home page shows the gate way to the authentic access to the digitalized responsive logical interface application.

**Figure 3:  Staff Login Module**

Figure 3 is the platform to which access are made to the system. In order to safeguard the entering mood, the system utilizes the token method of data transaction to grant access to legitimate users. The request is thereafter sent to the email (intranet) of the possible users. The code generated is per transaction. This allow the rightful users have full control of what happen within the system and automatically discourage who so ever that hope to access the system with bad intentions

**Figure 4: Registration Page**

Figure 4 enable new users to be added to the system. The data collected here is stored in the system database which only the administrator has access to. The administrator is saddled with the responsibility of granting privileges to users.

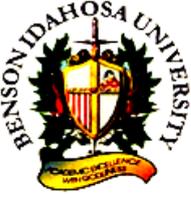

**Figure 5: Email Page**

To access the code (token) per transaction a user must have a pre-registered mail within the system module and through this page access is made to the code send to the email which is further retrieve to access the application system via the interface login.

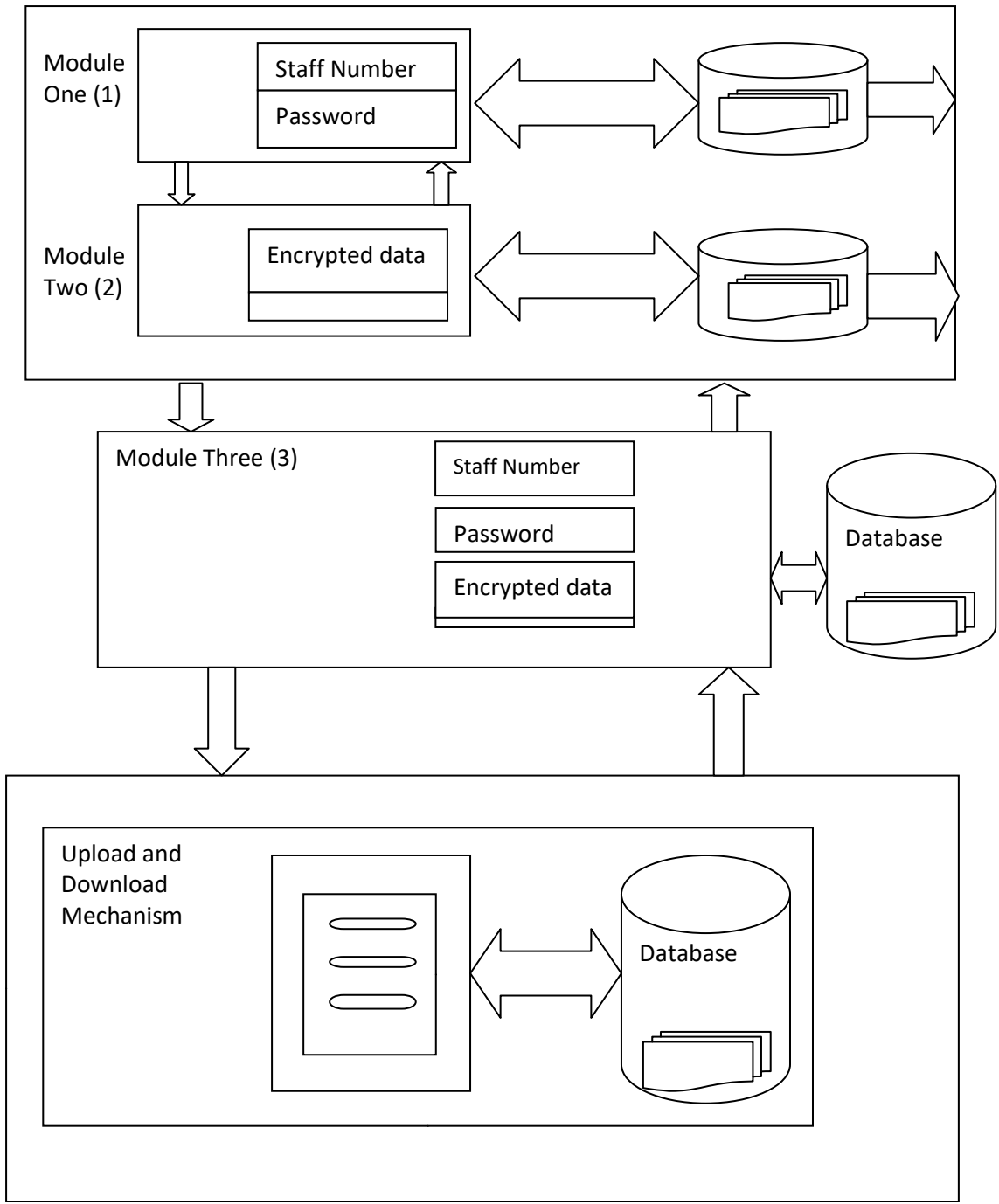

**Figure 6: Digitalized Responsive Logical Interface Application Architecture**

Figure 6 shows the structure of all components of the program designed. The first phase represented by module 1 and 2 represent the front end of the system where all necessary attributes are required for proper authentication. Module 3 represents the systematic verification against module 1 and 2 before access is made to the application.

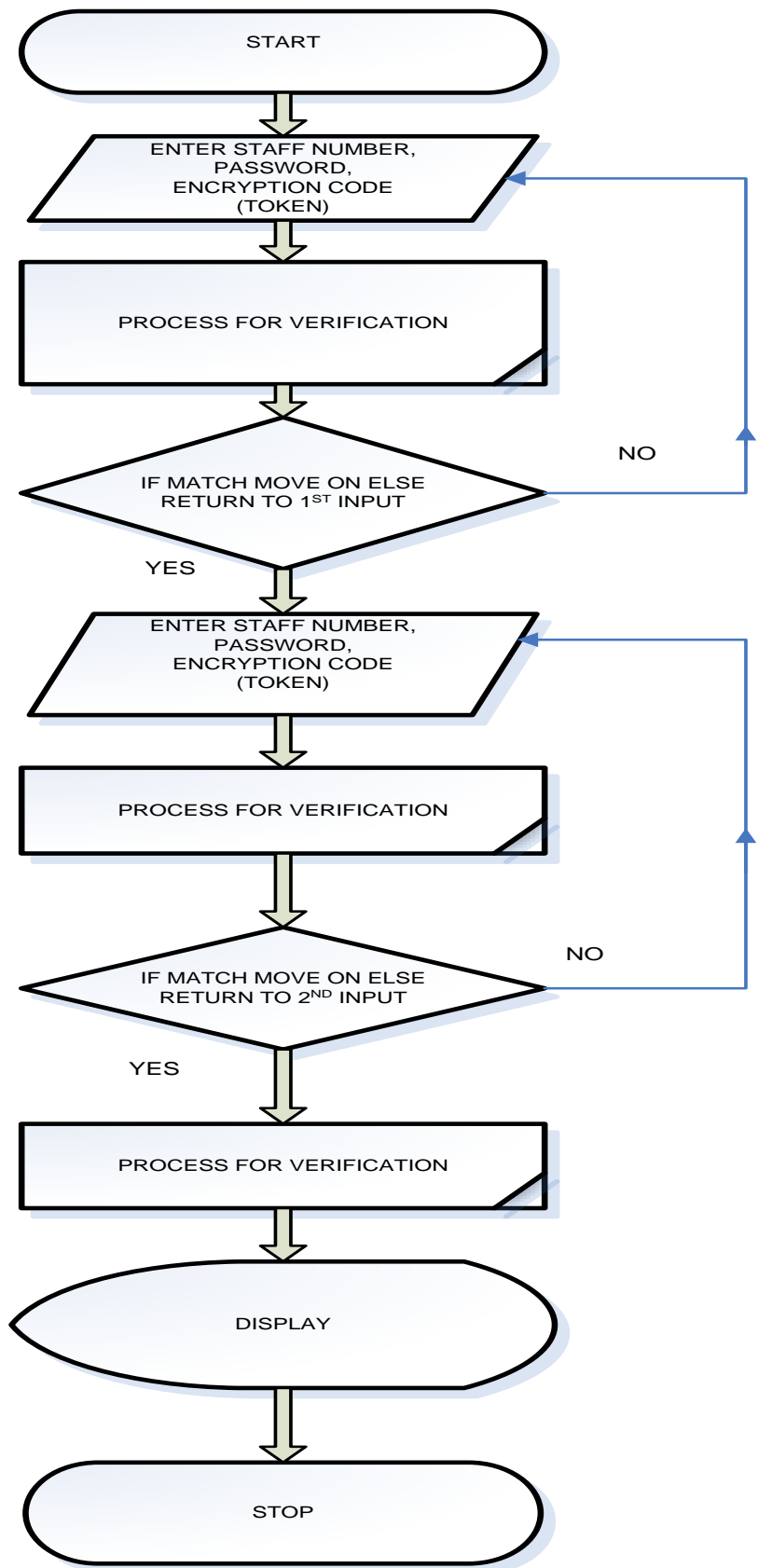

**Figure 7: Activity Diagram of Digitalized Responsive Logical Interface**

Figure 7 shows the navigation process of the digitalized responsive logical interface where all necessary decisions are put into scrutiny and thereafter feedback given as assigned to the designed program

## VI.  STATISTICAL VIEW

In this session, questionnaires were use to carry out need assessment through empirical knowledge from some notable tertiary institutions in Nigeria. All responses gotten were subjected to statistical analysis with the frequency adequately distributed and the percentage spelt out as seen in Table 1 and 2.

Table 3 shows the mean distribution against the agreed mean (3.00) and the standard deviation (SD) using the following formulas $mean\ (x) = \frac{\sum fx}{\sum f(n)}$ that is, where $n_1 + n_2 + \ldots + n_n = n$

$$Standard\ deviation\ (s) = \frac{\sqrt{\sum x^2 - (\sum X)2/n}}{n - 1}$$

**Table 1:  Attribute of people responses**

|  | Frequency | Percentage |
|---|---|---|
| (a) GENDER | | |
| Male | 26 | 65.0 |
| Female | 14 | 35.0 |
| (b) STAFF/STUDENT | | |
| Yes | 36 | 90.0 |
| No | 1.0 | 2.50 |
| Not Exactly | 3.0 | 7.50 |

(a) In gender, Table 1 shows that from the percentage the male within tertiary institutions are more aware of the operations of Examination and Record Unit.

(b) This part actually showed that 90% of the responses are either staff/student. This shows that the institution is making it much easier for adequate feedback

**Table 2: Basic knowledge of respondent on the operations of examination and record unit**

|  | Frequency | Percentage |
|---|---|---|
| (a) IS EXAMS AND RECORD PART OF REGISTRY | | |
| Yes | 33 | 82.5 |
| No | 4 | 10.0 |
| Not Exactly | 3 | 7.50 |
| (b) DOES THE UNIT HAVE COMPUTER SYSTEMS | | |
| Yes | 36 | 90.0 |
| No | 1 | 2.50 |
| Not Exactly | 3 | 7.50 |
| (c) IS THEIR OPERATIONS AUTOMATED | 4 | 10.0 |
| Yes | 25 | 62.5 |
| No | 11 | 27.5 |
| Not Exactly | | |
| (d) DOES THE SYSTEM USE USERNAME AND PASSWORD | | |
| Yes | 33 | 82.5 |
| No | 0 | 0.00 |
| Not Exactly | 7 | 17.5 |
| (e) IS THERE OTHER AUTHENTICATION METHODS | | |
| Yes | 2 | 5.00 |
| No | 27 | 67.5 |
| Not Exactly | 11 | 27.5 |

In terms of Table 2 (b) it was revealed that the unit have a computer system following the highest percentage of 90% and this seems to be a step in the right direction, meaning further operation can be achieve if the system is fully utilize  (c) with the 62.5% saying No, shows that the system use by examination and record unit is far from automated machine and with the current trend in data computing where sophisticated machines have taken over means the unit posses a feature danger to what is currently being practice (d) from the analysis, it was gathered that 82% (percent) agreed that the system currently being use by examination and record unit use Username and password (e) this part explained that apart from the username and password which is usually not adequate for data security, the system currently being use does not have other authentication modules for ensuring safety of data kept on the care of examination and record unit

**Table 3: Measuring the level of system awareness**

| S/N | Measuring the role of a Digitalized Responsive Logical Interface Application against the current practice in exams and record unit | Mean | Standard Deviation |
|---|---|---|---|
| 1 | Does password and user name has the capability of solidly protecting data from hackers | 2.3 | 1.1 |
| 2 | Do you think student's data are at risk due to poor authentication process | 4.38 | .797 |
| 3 | Do you support the idea of securing the data in the computer systems with structured components software | 4.63 | .533 |
| 4 | Do you support an idea to design a structure authenticated digitalized responsive logical interface application that has more security features for the computer systems and with an heterogeneous access *Agreed Mean (Mean > 3.00) | 4.65 | .6123 |

Table 3 shows the agreed mean, and standard deviation values gotten as a result of the statistical analysis ran on data gathered from this section.

$$mean\ (x) = \frac{\sum fx}{\sum f(n)}$$ that is, where $n_1 + n_2 + \ldots + n_n = n$

This section contained 4 questions measured on five point scale designed to measure the digitalized responsive logical interface application. The scale was scored as follows

Strongly Agree   -   5 points

Agree            -   4 points

Don't Know       -   3 points

Disagree         -   2 points

Strongly Disagree-   1 point

The agree mean = 3.00 is the benchmark for assessment and is compared against the mean in table 3. Apart from row one (1) mean which is 2.3 and showing that username and password is not solid enough for securing data which seems to be close to be truth, every other rows such as two (2), three (3) and four (4) have their mean higher than the benchmark mean. Meaning there is an agreement with the reality of the modern day computing or data handling and thereby giving support to the design and development of a digitalized responsive logical interface application capable of solving the problem associated with the old system.

***Testing the Hypothesis:***

1= SA+A = (28 + 11) = 39,     2=SD+D = (1+0) = 1.0

$$X^2 = \sum \frac{(f_o - f_e)^2}{f_e}$$

$f_o$ = Observe frequency in single category

$f_e$ = Expected, theoretical or hypothetical frequency

$\sum$ = Sum of

From the formula it is evident that the Chi-square is an index of the divergence of fact from hypothesis.

**Table 4: Hypothesis Analysis**

| Responses | 1 | 2 | |
|---|---|---|---|
| Observed ($f_0$) | 39 | 1.0 | 40 |
| Expected ($f_e$) | 20 | 20 | 40 |
| ($f_o$-$f_e$) | 19 | -19 | |
| ($f_o$-$f_e$)$^2$ | 361 | 361 | |
| ($f_o$-$f_e$)$^2$/$f_e$ | 18.05 | 18.05 | |

(Equal division of 40 persons in 2 cells)

$H_{o\ yes}$: $f_1 = f_2$

$H_{1,\ yes}$: $f_1 \neq f_2$

Decision Rule: Given α = .05: and 2 x 2 contingency table with degree of freedom (df)

= (r – 1) (c – 1) = (2- 1) (2-1) = the hypothesis has just a degree of freedom

If $X^2_{obs}$ > 5.99, Reject $H_0$

If $X^2_{obs}$ < 5.99, Accept $H_0$

$X^2 = \sum \dfrac{(f_o - f_e)^2}{f_e}$     18.05+ 18.05 = 36.1

The calculated value is greater than 5.99 i.e. $X^2$ which is 36.1 > 5.99, therefore reject $H_0$ accept $H_1$. From the calculation above the support for the idea to design and develop a well structure Digitalized Responsive Logical Interface Application is pretty high and demanding. This follows the setback currently associated with the mode of operation in Examinations and Records Unit in Nigeria tertiary institutions.

## VII.  CONCLUSION

The conceptual framework of a more robust, protected and all round access automated system for Examination and Record Unit for all Nigeria tertiary institutions  has not only become a reality by the design of this Digitalized Responsive Logical Interface Application but has also open doors for both homogeneous and heterogonous computing, staff of Examination and Record Unit of all tertiary institutions are no longer restricted to the four walls of their offices but can now easily and profoundly executes any operation anywhere using Digitalized Responsive Logical Interface Application. This will not only improve productivity among operational staff but also will increase their skill in computer proficiency. The Digitalized Responsive Logical Interface Application is an embedded security platform which measure the amount of individual that have legitimacy to the application. As stated above, the application captured both time and details of the person suing the system at a particular point in time so as to safeguard the data available to the unit. We are using this medium to make open to all researchers to improve on the area of biometric collaboration which was not cover